# Directional Acoustic Luneburg Lens Waveguide

**Liuxian Zhao[1], Timothy Horiuchi[1,2], Miao Yu[1,3]**

[1]Institute for Systems Research, University of Maryland, College Park, MD, 20742, USA

[2]Department of Electrical and Computer Engineering, University of Maryland, College Park, Maryland 20742, USA

[3]Department of Mechanical Engineering, University of Maryland, College Park, Maryland 20742, USA

lzhao128@umd.edu, timmer@umd.edu, mmyu@umd.edu

**ABSTRACT**

This paper investigates the acoustic Luneburg Lens (ALL) as a design framework for guiding acoustic wave propagation. In this study, we propose to develop an acoustic waveguide based on the characteristics of both acoustic wave focusing and collimation of cascaded ALLs. The continuous variation of refractive index of the ALL is achieved by using lattice unit cells with a graded filling ratio. A cascaded ALL waveguide device is fabricated based on the additive manufacturing technique. The experimental results obtained with this device are consistent with the numerical simulations and theoretical calculations.

1. Introduction

Advances in the development of acoustic metamaterials in recent years have enabled the realization of novel acoustic functionalities, such as acoustic absorption [1-3], acoustic cloaking [4, 5], and acoustic wave trapping [6, 7], acoustic focusing [8, 9], energy harvesting [10-12]. In addition, there is a growing interest in the study of acoustic waveguides based on acoustic metamaterials [13, 14]. Most existing acoustic metamaterial waveguides are based on tailoring of unit cells to achieve higher refractive indices or bandgaps. For example, Zhang *et al.* [15] developed an acoustic metamaterial waveguide based on acoustic analogies of topological insulators, in which the wave propagation was well confined to the desired trajectories. Zangeneh-Nejad *et al.* [16] proposed an acoustic analogue of high-index optical waveguides, which can be employed to realize acoustic equivalents of dielectric slab or strip waveguides. In addition, Otsuka *et al.* [17] investigated a waveguide mechanism based on phononic band-gap structures, in which rows of defects were intentionally designed in phononic crystals to effectively guide sound waves. Another method of achieving waveguides is based on cascaded Luneburg lenses. Mattheakis *et al.* [18] proposed a theoretical design of an optical waveguide based on cascaded Luneburg lenses, which can be used for sensing and nonlinear optics applications. This type of waveguides was later experimentally realized with periodic focusing of optical energy in a device by Smolyaninova *et al* [19]. More recently, the concept of optical waveguides based on cascaded Luneburg lens was applied for structural wave manipulation. Zhao *et al* [20] numerically and experimentally demonstrated cascaded structural Luneburg lens for guiding elastic wave propagations in thin plate structures.

Inspired by the previous studies of cascaded Luneburg lens for guiding the propagation of optical waves and structural waves, in this work, we propose to use cascaded Acoustic Luneburg lens (ALL) for guiding acoustic wave propagation. Similar to its counterpart of optical and structural Luneburg lenses, ALL has a gradient refractive index profile that increases radially from the outer surface of the lens to its centre. There are two characteristics of the ALL: wave collimation and focusing, both of which have been explored [21] [22]. The proposed cascaded ALL waveguide is based on the combination of these two characteristics. When a point source is used for excitation, an ALL can convert the point source into a plane wave, and then the neighbouring lens will focus the plane wave perfectly at a point on the other side of the lens. In this way, the source energy can be guided to propagate along the desired direction, as shown in Figure 1, which indicates that the cascaded ALLs can serve as a waveguide for acoustic waves.

Although the collimation and focusing properties of acoustic Luneburg lens have been studied in the literature, this work proposes a novel application of these properties for achieving acoustic waveguides. In recent searches of the acoustics literature, studies of acoustic waveguides based on cascaded acoustic Luneburg lenses were not apparent. One potential application of the acoustic Luneburg waveguide is on-chip waveguide as part of an acoustic circuit for acoustic communications [23]. At the designed working frequency of 40 kHz, the acoustic Luneburg waveguide can work as a passive acoustic filter while guiding the wave propagation along a designed path. At other frequencies, the acoustic wave will diverge. In this case, using acoustic Luneburg waveguide provides clear advantages of avoiding the use of complex and expensive electronic devices.

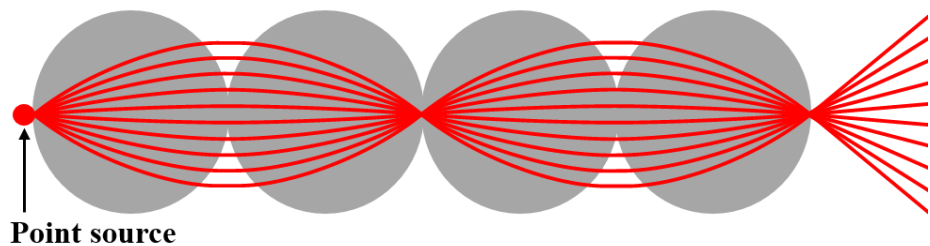

**Point source**

**Figure 1: Schematic of acoustic waveguide with cascaded Luneburg lens.**

**2. Cascaded Acoustic Luneburg Lens Design**

As stated previously, the ALL is a spherically symmetric gradient-index lens. The lens is defined by a circular region with $r \leq R$, where $r$ is the radial distance from the lens center and $R$ is the radius of the lens. Based on the principle of optical Luneburg lens [24, 25], the radially gradient refractive index satisfies the following governing equation:

$$n(r) = \sqrt{2 - (r/R)^2}, \qquad (1)$$

The ray trajectories of acoustic wave are calculated based on the ray tracing technique. The governing equations of the ray tracing technique for a spherical symmetric lens are given in the following equations [26, 27]:

$$\begin{cases} \frac{dr}{d\theta} = r \frac{1}{\tan(\alpha)} \\ \frac{d\alpha}{d\theta} = -1 - \frac{r}{n} \frac{dn}{dr} \end{cases} \qquad (2)$$

where $r$ and $\theta$ are the radial and the angular coordinates in the polar coordinate system, respectively, and $\alpha$ is the angle between the wave vector $k$ and the radial coordinate $r$.

For proof-of-concept, we consider a perfect ALL with a radius $R = 0.02$ m. Based on Equation (1), the distribution of the refractive index of the ALL is plotted in Figure 2 (a) and (b). In this study, the ray tracing method is used to illustrate acoustic wave collimation and focusing properties of the ALL. When a 40 kHz point source located at $(x, y) = (-R, 0)$ is used for excitation, collimated waves can be generated, as shown in Figure 2(c). On the contrary, the ray trajectories of acoustic wave focusing are shown in Figure 2(d), when a family of incident rays propagate through the lens from $-x$ toward $+x$, with the excitation source located at $x = -1.2R$ and $y = -R$ to $R$.

Based on Figure 2(a), the refractive index profile of the perfect ALL has a range between 1 and 1.42. Based on our prior work [21], we designed and fabricated a 2D ALL with a radius of $R = 0.02$ m at a working frequency of 40 kHz. We used a 3D lattice truss consisting of three orthogonal beams as a unit cell, as shown in Figure 2(e). This unit cell allows the construction of a stable three dimensional ALL device. In principle, a range of variable refractive index values can be obtained using this unit cell via changing the filling ratio. In our design, the length and width of the beam are $a_0d$, where $d$ is the periodicity of the unit cell. In the following studies, we chose $d = 2$ mm. The factor $a_0$ can be tailored to change the filling ratio of the unit cell, and hence change the refractive index correspondingly. The standard retrieval technique was used to obtain the refractive index of the each unit cell [28]. The graded refractive indices along the radial distance were selected (the red dots in Figure 2(b)). In this study, 5 layers of unit cells were stacked in the thickness direction to form the 2D ALL device, as shown in Figure 2(f).

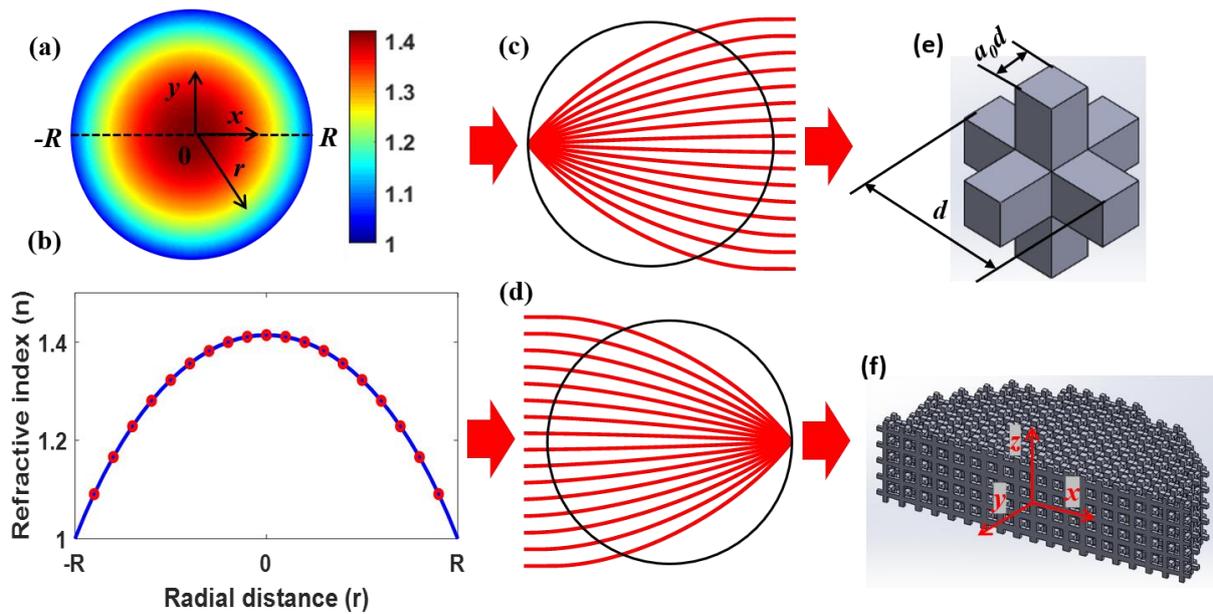

**Figure 2: (a) Refractive index distribution of ALL. (b) Refractive index profile (blue line) along the radial distance (dashed line in (a)), and the red dots indicate the selected refractive indices for the designed ALL. (c) and (d) Ray trajectories for acoustic collimation and focusing based on Luneburg lens. (e) Unit cell used to design ALL. (f) Cross section of 2D circular ALL.**

## 3. Numerical Modelling and Experimental Results

In this study, we explored the capability of the cascaded ALL for guiding of acoustic waves. Full 3D wave simulations were conducted by using commercial software COMSOL. The 2D lens was built based on lattice unit cells with graded refractive indices. The entire air domain in the simulation were $l_0 \times w_0 \times h_0 = 0.17$ m $\times 0.04$ m $\times 0.01$ m. The cascaded ALL was constructed with four ALLs with a starting point at $(x, y) = (0$ m, $0$ m$)$. The ALLs were simulated by using hard boundary conditions. The radiation boundary conditions were applied on the outer boundaries to assume infinite air space. A point source located at $(x, y, z) = (0$ m, $0$ m, $0$ m$)$ was used for excitation at the frequency of 40 kHz, as shown in Figure 3(a). The simulation results were obtained from the central surface ($z = 0$ m) of the five layers, and the acoustic pressures were collected along the dashed line (located at $x = 0.16$ m, $z = 0$ m, and $y = -R$ to $R$).

Experiments were also carried out in an anechoic chamber to validate the simulation results. The 2D ALLs were fabricated with the same dimensions as the numerical models, as shown in Figure 3(b). The fabrication was based on the additive manufacturing technique by using a 3D printer (Stratasys Objet500 Connex3) with a resolution of up to 200 microns. A high strength material (Clear Resin) was used with Young's modulus $E=1.6$GPa, density $\rho= 1170$ g/m$^3$, and Poisson's ratio $v=0.34$. Figure 3(b) shows the experimental setup for characterizing the ALL waveguide. The four ALLs were fixed on a supporting structure consecutively. A miniature speaker (MA40S4S from Murata Manufacturing Co., Ltd.) with a diameter of 9 mm was placed next to the first lens and used as a point source for excitation of acoustic waves of a central frequency of 40 kHz, with the input voltage amplitude $V_p = 5$ V. The acoustic field distribution along the same line as in the simulation was collected by using a fiber optic probe [29]. A translational stage was used to facilitate the measurement of the acoustic field distribution along the $y$ direction.

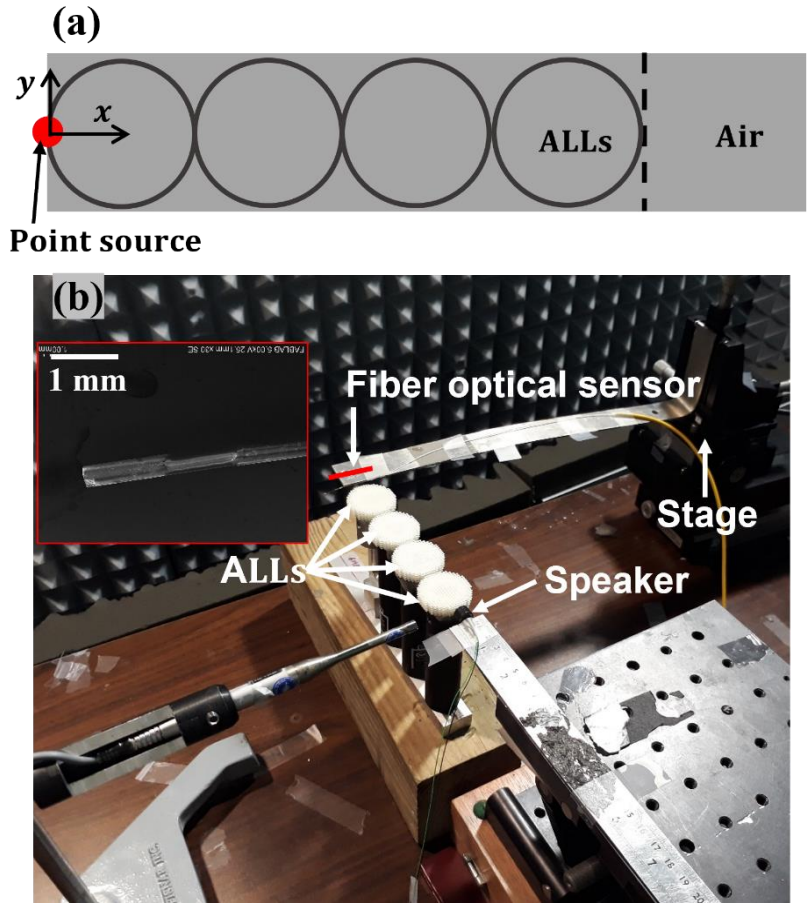

**Figure 3: (a) Schematic of numerical model with four cascaded ALLs as waveguide (top view) and (b) Experimental setup for characterization of the cascaded ALL waveguide. The red line indicates the location of the fiber optic sensor, and the inset shows the Scanning Electron Microscope (SEM) image of the fiber optic sensor.**

By using the above mentioned analytical and numerical methods, the cascaded ALL based waveguide was explored. The ray trajectories of acoustic wave propagation and interaction were obtained, as shown in Figure 4 (a). A point source on the edge of the first ALL excited a family of incident rays, which travelled through four cascaded ALLs and experienced alternating collimation and focusing before exiting the waveguide with circular waveforms. In addition, numerical simulations were obtained and overlayed with the ray trajectories, as shown in Figure 4 (a). A good agreement between analytical and numerical results can be observed, which proves that the cascaded ALL can serve as an acoustic waveguide.

Furthermore, the acoustic pressure amplitude was collected from numerical simulations along the dashed lines in Figure 4 (a). At the same time, the acoustic pressure amplitude was calculated numerically for a reference model without the cascaded ALL. A normalized amplitude is defined as $A_N = \{A_{w/}, A_{w/o}\}/\max(A_{w/})$, where $A_{w/}$ and $A_{w/o}$ indicate the acoustic wave amplitude for the case with and without the cascaded ALL, respectively. As shown in Figure 4(b), the acoustic wave amplitude was increased more than 4 times by using the cascaded ALL compared to the case without the ALL.

In addition, experimental studies were carried out to validate the cascaded ALL based wave waveguide. The scanning of acoustic pressure was performed along the same locations as that in the numerical simulations (dashed line in Figure 4(a)). Similarly, reference pressures without the cascaded ALL were also measured experimentally. The experimental results of the normalized acoustic pressure are presented in Figure 4(c). The experimental results are consistent with the numerical simulations, which demonstrate an increase of acoustic amplitude up to 3.2 times by using the cascaded ALL.

Furthermore, the transmission loss between two consecutive Luneburg lenses was investigated. The acoustic pressure values ($P_A$ and $P_B$) at points A and B shown in Figure 4(a) were obtained in both numerical simulations and experiment. The transmission loss (*TL*) was calculated as $L = 10 log_{10}(\frac{P_A^2}{P_B^2})$. The results show that the transmission loss between the two consecutive Luneburg lenses is 1.36 dB in the numerical simulations and 1.97 dB in the experiment; while without the Luneburg lens, the transmission loss is 7.84 dB in the numerical simulations and 9.05 dB in the experiment. These results indicate that the cascaded ALLs exhibit a good performance for guiding acoustic wave propagation with a relatively low loss.

To analyze the effect of the speaker size, numerical simulations were performed to compare the performance of the ALL waveguide by using the 9 mm-speaker excitation and the point source excitation. The obtained transmission losses between two consecutive Luneburg lenses (Points A and B) are 1.36 dB and 1.49 dB for the point source excitation and the 9 mm-speaker excitation, respectively. These results indicate that the effect of speaker size is negligible.

The smaller gain obtained in experiment compared with numerical simulations are due to the following reasons: (1) the additive manufacturing technique inevitably results in breaking truss unit cells after post processing, which can influence the performance of acoustic transmission; (2) there might be slight misalignments for the source location and measurement positions, which can result in some measurement errors. Nevertheless, these experimental results successfully demonstrated the wave guiding capacities of the cascaded ALL.

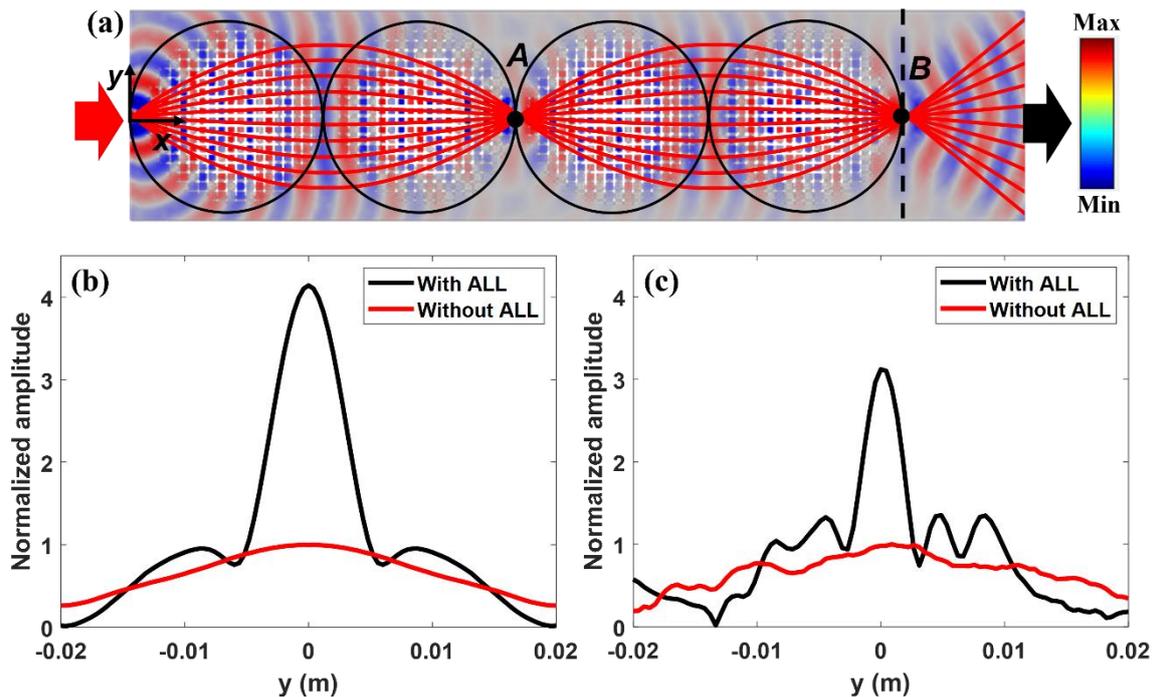

**Figure 4: Characterization of cascaded ALL waveguide. (a) Numerical simulation results of acoustic pressure distribution overlaid with ray trajectories. Comparison of acoustic amplitude along the dashed line in (a) between the cases with and without ALLs in (b) numerical simulations and (c) experiments.**

## 4. Conclusions

We analytically, numerically, and experimentally demonstrated for the first time that acoustic waveguide can be achieved by using the cascaded ALLs. The ALL used in the waveguide is based on variation of the filling ratio of lattice unit cells, which renders a graded change of refractive index along the radial direction. This ALL allows for both acoustic wave focusing and collimation. By taking advantage of the properties of the ALL, an acoustic waveguide was realized based on cascading four ALLs. The ray trajectory method was used to calculate the wave propagation through the acoustic wave waveguide, demonstrating its working principles. Furthermore, numerical simulations were carried out to characterize the performance of the acoustic wave waveguide, which were validated by the experimental studies. In both the numerical and experimental studies, the cascaded ALL waveguide demonstrated pressure gains of up to 4.1 and 3.2, respectively, compared with the cases without the ALL waveguide. This work provides a new solution for the development of acoustic waveguides.

**Conflict of Interest**

The authors declare no conflict of interest.

**Acknowledgements**

This work was supported by the AFOSR Center of Excellence on Nature-Inspired Flight Technologies and Ideas and USDA NIFA Sustainable Agricultural Systems (award number 20206801231805).

**Dada Availability**

The data that support the findings of this study are available from the corresponding authors upon reasonable request.